\documentclass{new_tlp}
\usepackage{times}
\usepackage{helvet}
\usepackage{courier}
\usepackage{graphicx}
\usepackage{enumitem}
\usepackage[cmtip,arrow]{xy}
\usepackage[curve]{xypic}
\usepackage[comments]{krudces}
\usepackage{txfonts}
\usepackage{lmodern}

\makeatletter
\renewcommand\paragraph{%
  \@startsection{paragraph}{4}{\z@}
    {-13\p@ \@plus -1.5\p@ \@minus -1.5\p@}
    {-0.5em}
    {\normalfont\normalsize\itshape\raggedright\bfseries}%
}
\makeatother

\renewcommand{\signature}{\tuple{\at,\lb}}
\renewcommand{\nphead}{l}
\renewcommand{\nnhead}{k}
\renewcommand{\npbody}{m}
\renewcommand{\nnbody}{n}

\newcommand{\noatomlabelss}{\lambda}

\newcommand{\noatomlabels}[1]{\noatomlabelss(#1)}

\newcommand{\warrow}{\ \leftsquigarrow \ }
\newcommand{\rchoice}[2]{\mathtt{ch}(#1,#2)}

\newcommand{\stmodels}{\models^{st}}
\newcommand{\lbr}{\lb_{\mathcal{R}}}

\newcommand{\infection}{\mathit{infection}}
\newcommand{\fever}{\mathit{fever}}
\newcommand{\dead}{\mathit{dead}}
\newcommand{\wet}{\mathit{wet}}
\newcommand{\shoot}{\mathit{shoot}}
\newcommand{\bulletproof}{\mathit{bulletproof}}
\newcommand{\ab}{\mathit{ab}}
\newcommand{\harvey}{\mathit{harvey}}

\newcommand{\billystab}{\mathit{joker\_stab}}
\newcommand{\heads}{\mathit{head}}
\newcommand{\tails}{\mathit{tails}}
\newcommand{\noheart}{\mathit{no\_heartbeat}}
\newcommand{\loaded}{\mathit{loaded}}

\newcommand{\prison}{\mathit{prison}}

\newcommand{\head}{\mathit{head}}
\newcommand{\body}{\mathit{body}}

\newcommand{\wleq}{\sqsubseteq} 
\newcommand{\wless}{\sqsubset} 
\newcommand{\Atoms}{\mathit{Atoms}} 

\title{Justifications for Programs with\\ Disjunctive and Causal-choice Rules\footnote{This research was partially supported by Spanish Project TIN2013-42149-P.} }

\author[P. Cabalar \& J. Fandinno ]
         {Pedro Cabalar and Jorge Fandinno\\
          Department of Computer Science\\
		  University of Corunna, Corunna, Spain\\
		  \email{\{cabalar, jorge.fandino\}@udc.es}
		 }

\newcommand{\veeh}{\vee}

\begin{document}

\setcounter{page}{1}
\label{firstpage}
\maketitle

\begin{abstract}
In this paper, we study an extension of the stable model semantics for disjunctive logic programs where each true atom in a model is associated with an algebraic expression (in terms of rule labels) that represents its justifications.
As in our previous work for non-disjunctive programs, these justifications are obtained in a purely semantic way, by algebraic operations (product, addition and application) on a lattice of causal values.
Our new definition extends the concept of \emph{causal stable model} to disjunctive logic programs and satisfies that each (standard) stable model corresponds to a disjoint class of causal stable models sharing the same truth assignments, but possibly varying the obtained explanations.
We provide a pair of illustrative examples showing the behaviour of the new semantics and discuss the need of introducing a new type of rule, which we call \emph{causal-choice}. 
This type of rule intuitively captures the idea of ``$A$ may cause $B$'' and, when causal information is disregarded, amounts to a usual choice rule under the standard stable model semantics.
\end{abstract}



\section{Introduction}
\label{sc:introduction}

The strong connection between Non-Monotonic Reasoning~(NMR) and Logic Programming~(LP) semantics for default negation has made possible that LP tools became nowadays an important paradigm for Knowledge Representation~(KR) and problem-solving in Artificial Intelligence~(AI). In particular, \emph{Answer Set Programming}~(ASP)  \cite{niemela1999,MT99} 
is an LP paradigm based on the \emph{stable models semantics}~\cite{GelfondL88,gelfond1991classical} that has become established as a preeminent framework for practical NMR with applications in diverse areas of AI including planning, reasoning about actions, diagnosis, abduction and beyond~\cite{baral2003knowledge}.
An important difference between most LP semantics (including the stable model semantics) and classical models (or even \review{R3.3}{some} other NMR approaches) is that true atoms in LP must be founded or justified by a given derivation.
These \emph{justifications} are not provided in the semantics itself, but can be syntactically built in some way in terms of the program rules, as studied in several approaches~
\cite{specht1993generating,pemmasani2004online,pontelli2009justifications,deneckerBS15,schulzT2016justifying}.

Rather than manipulating justifications as mere syntactic objects,
two recent approaches have considered extended multi-valued semantics LP extensions where justifications are treated as \emph{algebraic} constructions in terms of rule labels: \emph{Why-not Provenance}~\cite{damasio2013justifications} and~\emph{Causal Graph Justifications} (CJ)~\cite{CabalarFF14}.
For instance, as an illustration of the latter, consider the following positive program~\newprogram\label{prg:dead}
\begin{IEEEeqnarray}{l C ? c C l}
r_1 &:& \dead &\lparrow& \shoot
	\label{eq:r1}
\\
r_2 &:&\shoot &\lparrow& \harvey
	\label{eq:r2}
\\
	&& \harvey
	\label{eq:suzy}
\end{IEEEeqnarray}
\REVIEW{R3.4}%
\noindent
It has been showed in~\cite{CabalarFF14} that program~\programref{prg:dead} has a least causal model $I$, which not only makes $\dead$ true, but also assigns the expression $I(\dead) = \harvey \cdotl r_2 \cdotl r_1$ capturing the successive application of $r_2$ and then $r_1$.
\ENDREV
The availability of such information on justifications \emph{inside the models} of a program can be of great interest for Knowledge Representation (KR) since it allows defining new types of literals to inspect the causal relations among derived atoms.

For instance, \cite{fandinno2015aspocp} introduced constructs like:
\begin{IEEEeqnarray}{l C ' l C l}
&& \prison(\harvey) &\lparrow&  \harvey \sufficient \dead
	\label{eq:prison}
\end{IEEEeqnarray}
\noindent where `$\sufficienttt$' decides whether $\harvey$ has been a sufficient cause for $\dead$ by completely relying on the semantic definitions, rather than making a syntactic analysis of the program.
Note that avoiding the latter is crucial to achieve an elaboration tolerant representation, since causation may be affected by multiple levels of indirect effects or the interplay with defaults like inertia.
\cite{Fandinno16} extends this formalism allowing the use of `$\necessarytt$' and `$\contributedtt$' to capture whether $\harvey$ has been necessary for (resp. contributed to) $\dead$.
To the best of our knowledge, no other formalism integrates this kind of literals in LP or allows deriving new conclusions from information about cause-effect relations.
Unfortunately, CJ is only defined for non-disjunctive logic programs -- this holds, in fact, for all the approaches aforementioned.
This is obviously a drawback from an ASP point of view, since no CJ justifications were defined for programs with disjunction, but also, from a KR point of view, since disjunction is a natural way for expressing non-deterministic causation.

In this paper, we investigate an extension of the CJ~semantics for disjunctive logic programs, in which, per each standard stable model, we will obtain a class of causal stable models providing explanations for the derived (true) atoms.
As an example, let program~\newprogram\label{prg:coin} be the result of replacing~\eqref{eq:r2} in~\programref{prg:dead} by the following rules
\begin{IEEEeqnarray}{l C ? c C l}
r_2 &:& \shoot &\lparrow& \tails
	\label{eq:r2b}
	\\
r_3 &:& \heads \veeh \tails &\lparrow& \harvey
	\label{eq:r3}
\end{IEEEeqnarray}
stating that Harvey throws a coin and only shoots when he gets tails.
This program has two standard stable models: $\set{\harvey, \heads}$ and $\set{\harvey, \tails, \shoot, \dead}$.
In the latter, the explanation for $\dead$ will have the form
\begin{gather}
\harvey \cdotl r_3^{\tails} \cdotl r_2 \cdotl r_1
	\label{eq:introduction.term}
\end{gather}
where $r_3^{\tails}$ points out that the disjunct $\tails$ in $r_3$ has been effectively applied.

We summarise our contributions as follows.
\begin{itemize}
\item We define a multi-valued semantics for disjunctive logic programs\footnote{For simplicity, in this paper, we will not consider causal literals as in rule~\eqref{eq:prison}, but the extension can be done in a similar way as~\cite{fandinno2015aspocp} extended~\cite{CabalarFF14}.} in which each atom~$\rA$ is associated to an algebraic expression formed by rule labels and three operations:  a product `$*$' representing conjunction or joint causation; an addition~`$+$' separating alternative causes; and a non-commutative product `$\cdot$' that stands for rule application.
We show that there is a one-to-one correspondence between the standard stable models and a corresponding class of \emph{causal stable models} that provide explanations for each true atom in each model (Section~\ref{sec:csm}).

\item We show that disjunctive rules are not enough to capture informal statements like ``$A$ may cause $B$'' and introduce the notion of \emph{causal-choice} rule to represent this kind of statement.
When causal information is disregarded, causal-choice rules
amount to \review{R3.5}{usual choice rules} under the standard stable model semantics (Section~\ref{sec:causal-choice.rules}). 
\end{itemize}
The rest of the paper is organised as follows.
The next section revises the standard stable model semantics for disjunctive logic programs and  recalls the causal algebra introduced in~\cite{CabalarFF14}.
Section~\ref{sec:csm} provides the new definition of causal stable models for disjunctive programs and shows some correspondence properties with standard stable models.
In Section~\ref{sec:causal-choice.rules}, we introduce causal-choice rules, explain their behaviour with an example and, again, provide a correspondence with standard ASP. 
Section~\ref{sec:related} revises some related approaches and, finally, Section~\ref{sec:conc} concludes the paper.
Proofs of formal results from the paper can be found in an extended version~\cite{CabalarF16arXiv}.

\vspace{-0.20cm}
\section{Preliminaries}
\label{sec:causal.semantics}

We start by recalling some standard LP definitions, with the only extension of rule labels. A \emph{signature} is a pair \signature\ of sets that respectively represent a set of \emph{atoms} (or \emph{propositions}) and a set of \emph{labels}. As usual, a \emph{literal} is defined as an atom $\rA$ (positive literal) or its default negation $\Not \rA$ (negative literal).

\begin{definition}[Program]\label{def:program}
Given a signature $\signature$, a \emph{rule} is an expression of the form
\begin{eqnarray}
r_i: \ \ \rH_1 \veeh \dotsc \veeh \rH_\nphead
 \veeh \Not \rH_{\nphead + 1} \veeh \dotsc \veeh \Not \rH_\nnhead
		\ \lparrow \ \rB_1, \dotsc, \rB_\npbody,
		\Not \rB_{\npbody+1}, \dotsc, \Not \rB_\nnbody
    \label{eq:rule}
\end{eqnarray}
where \mbox{$r_i\in \lb$} is a label and all $\rA_j \in \at$ and $\rB_j$ are atoms \review{R3.6}{with $0 \leq \nphead \leq \nnhead$ and $0\leq \npbody \leq \nnbody$.}
A \emph{program}~$\cP$ is a set of rules.\qed
\end{definition}

Given a rule~$\R$ of the form of~\eqref{eq:rule},
by $\head^+\!(\R)\eqdef\set{\rH_1 \dotsc  \rH_\nphead}$ and
$\head^-\!(\R)\eqdef\set{\rH_{\nphead +1}  \dotsc \rH_\nnhead}$,
we respectively denote the positive and negative head of~$\R$.
By $\head(\R)\eqdef \head^+(\R) \cup \Not\head^-(\R)$,
 we denote the whole head of~$\R$ where $\Not \head^-(\R)$ stands for the set $\setm{ \Not \rA }{ \rA \in \head^+(\R) }$.
By
$\body^+(\R)\eqdef\set{\rB_1 , \dotsc , \rB_\npbody}$ and
$\body^-(\R)\eqdef\set{\rB_{\npbody+1} , \dotsc , \rB_\nnbody}$,
we respectively denote the positive and negative body of~$\R$
and, by
$\body(\R)\eqdef \body^+(\R) \cup \Not \body^-(\R)$,
its whole body.
A rule $\R$ is said to be \emph{normal} iff its head is a single positive literal, that is,
$\head(\R)=\head^+(\R)=\set{\rA}$ and it is said to be
\emph{positive} iff all literals in it are positive,
that is,
$\head^-(\R)=\body^-(\R)=\emptyset$.
A normal rule with empty body, $\body(\R)=\emptyset$, is called a \emph{fact} and we usually represent it omitting the body and sometimes the symbol~`$\leftarrow$.'
Furthermore, we \REVIEW{R2.1}will also allow atom names to be used as labels, that is, $\at \subseteq \lb$,
\ENDREV
 and that fact $\rA$ in a program actually stands for the labelled rule~$(\rA : \rA \leftarrow)$.
Any rule that is not a fact has a label from $\lbr \eqdef \lb \backslash \at$.
When $\head(\R) = \emptyset$, we say that the rule $\R$ is a \emph{constraint} and we may omit the rule label.
A program $P$ is said to be \emph{normal} or \emph{positive} when all its rules are normal or positive, respectively.

We say that a set of atoms $S \subseteq \at$ 
is \emph{closed} under $P$ if and only if, for every rule $\R \in P$,
 $\head^+(\R) \cap S \neq \emptyset$ 
whenever $\body^+(\R) \cup \head^-(\R) \subseteq S$ and
$\body^-(\R) \cap S = \emptyset$.
We recall next the definitions of reduct and stable model.
The \emph{reduct} of a program $P$ w.r.t. a set of atoms $S \subseteq \at$, in symbols $P^S$, is the result of
\vspace{-0.15cm}
\begin{enumerate}
\item removing all rules such that $\rB \in S$ for some $\rB \in \body^-(\R)$,
\item removing all rules such that $\rB \notin S$ for some $\rB \in \head^-(\R)$,
\item removing all the negative literals for the remaining rules.
\end{enumerate}
\vspace{-0.15cm}

\noindent
A set of atoms $S \subseteq \at$ is a \emph{GL-stable model} of a program~$P$ iff it is a \mbox{$\subseteq$-minimal} closed set under the positive program~$P^S$.

We will also need the following definitions from~\cite{CabalarFF14}.

\begin{definition}[Term]
Given a set of labels $Lb$, a \emph{term} $t$ is recursively defined as one of the following expressions
\begin{gather*}
t \ \ ::= \ \ l \ \ \Big| \ \ \prod S \ \ \Big| \ \ \sum S \ \ \Big| \ \ t_1 \cdot t_2
\end{gather*}
where $l \in Lb$ is a label, $t_1, t_2$ are in their turn terms and $S$ is a (possibly empty and \review{R3.7}{possibly infinite}) set of terms.\QED
\end{definition}

\noindent
\REVIEW{R3.8}
When $S = \set{t_1, \dotsc, t_n}$ is a finite set, we will write $t_1 * \dotsc * t_n$ and $t_1 + \dotsc + t_n$ instead of~$\prod S$ and $\sum S$, respectively. 
The empty sum and empty product are respectively represented as $0$ and $1$.
We assume that application~`$\cdot$' has higher priority than product~`$*$' and, in its turn, product~`$*$' has higher priority than addition~`$+$'.
\ENDREV
As commented in the introduction, product~`$*$' represents join causation.
Consider for instance a program~\newprogram\label{prg:loaded} obtained by adding the fact $loaded$ to program~\programref{prg:dead} and replacing rule~\eqref{eq:r1} by
\begin{IEEEeqnarray}{l C ? r C l}
r_1 &:& \dead &\lparrow& \shoot, \loaded
	\label{eq:r1.loaded}
\end{IEEEeqnarray}
\REVIEW{R3.9}%
Program~\programref{prg:loaded} will assign the expression
$(\loaded * \harvey \cdotl r_2) \cdotl r_1$ to the atom $\dead$.
On the other hand, addition~`$+$' is used to capture alternative causes.
As an illustration, if we add the rules:
\vspace{-0.4cm}
\begin{IEEEeqnarray}{l C ? r C l}
 & & & &\billystab 
	\label{eq:r4a}\\
r_4 &:& \dead &\lparrow& \billystab
	\label{eq:r4}
\end{IEEEeqnarray}
to program~\programref{prg:dead}, meaning that the Joker stabs the victim, apart from the simultaneous Harvey's shot,
then this new program~\newprogram\label{prg:billy} will assign the expression
$(\loaded * \harvey \cdotl r_2) \cdotl r_1 + \billystab \cdotl r_4$ reflecting the existence of two alternative and independent causes.
\ENDREV
\begin{figure}[htbp]
\begin{center}
\footnotesize
\newcommand{\titleSep}{0pt}
\newcommand{\contentSep}{-10pt}
\newcommand{\rowSep}{5pt}
$
\begin{array}{c}
\hbox{\em Associativity}\vspace{\titleSep}\\
\hline\vspace{\contentSep}\\
\begin{array}{r@{\ }c@{\ }r@{}c@{}l c r@{}c@{}l@{\ }c@{\ }l@{\ }}
t & \cdot & (u & \cdot & w) & = & (t & \cdot & u) & \cdot & w\\
\\
\end{array}
\end{array}
$
\ \
$
\begin{array}{c}
\hbox{\em Absorption}\vspace{\titleSep}\\
\hline\vspace{\contentSep}\\
\begin{array}{r@{\ }c@{\ }c@{\ }c@{\ }l c r@{\ }c@{\ }r@{\ }c@{\ }c@{\ }c@{\ }c@{\ }l@{\ }}
&& t &&& = & t & + & u & \cdot & t & \cdot & w \\
u & \cdot & t & \cdot & w & = & t & * & u & \cdot & t & \cdot & w
\end{array}
\end{array}
$
\ \
$
\begin{array}{c}
\hbox{\em Identity}\vspace{\titleSep}\\
\hline\vspace{\contentSep}\\
\begin{array}{rc r@{\ }c@{\ }l@{\ }}
t & = & 1 & \cdot & t\\
t & = & t & \cdot & 1
\end{array}
\end{array}
$
\ \
$
\begin{array}{c}
\hbox{\em Annihilator}\vspace{\titleSep}\\
\hline\vspace{\contentSep}\\
\begin{array}{rc r@{\ }c@{\ }l@{\ }}
0 & = & t & \cdot & 0\\
0 & = & 0 & \cdot & t\\
\end{array}
\end{array}
$
\\
\vspace{\rowSep}
$
\begin{array}{c}
\hbox{\em Indempotence}\vspace{\titleSep}\\
\hline\vspace{\contentSep}\\
\begin{array}{r@{\ }c@{\ }l@{\ }c@{\ }l }
l & \cdot & l  & = & l\\
\\
\\
\end{array}
\end{array}
$
\hspace{.05cm}
$
\begin{array}{c}
\hbox{\em Addition\ distributivity}\vspace{\titleSep}\\
\hline\vspace{\contentSep}\\
\begin{array}{r@{\ }c@{\ }r@{}c@{}l c r@{}c@{}l@{\ }c@{\ }r@{}c@{}l@{}}
t & \cdot & (u & + & w) & = & (t & \cdot & u) & + & (t & \cdot & w)\\
( t & + & u ) & \cdot & w & = & (t & \cdot & w) & + & (u & \cdot & w)\\ \\
\end{array}
\end{array}
$
\hspace{.05cm}
$
\begin{array}{c}
\hbox{\em Product\ distributivity}\vspace{\titleSep}\\
\hline\vspace{\contentSep}\\
\begin{array}{rcl}
c \cdot d \cdot e & = & (c \cdot d) * (d \cdot e) \ \hbox{with} \ d \neq 1 \\
c \cdot (d*e)     & = & (c \cdot d) * (c \cdot e) \\
(c*d) \cdot e     & = & (c \cdot e) * (d \cdot e)
\end{array}
\end{array}
$
\end{center}
\vspace{-5pt}
\caption{Properties of the  `$\cdot$'operators:
$t,u,w$ are terms, $l$ is a label and $c,d,e$ are terms without addition~`$+$'.
Addition and product distributivity are also satisfied over infinite sums and products.}
\label{fig:appl}
\end{figure}
\begin{definition}[Value]
\label{def:values}
\emph{(Causal) values} are the equivalence classes of terms under axioms for a completely distributive (complete) lattice with meet~`$*$' and join~`$+$' plus the axioms of Figure~\ref{fig:appl}.
The set of values is denoted by~$\values$.
Furthermore, by $\causes$ we denote the subset of causal values with some representative term without addition~`$+$'.\QED
\end{definition}

All three operations, `$*$', `$+$' and `$\cdot$' are associative. Product~`$*$' and addition `$+$' are also commutative, and they satisfy the usual absorption and distributive laws with respect to infinite sums and products of a completely distributive lattice. 
The lattice order relation is defined as:
\begin{IEEEeqnarray*}{c"C"c"C"c}
t \leq u & \text{ iff } & t * u = t & \text{ iff } & t + u = u
\end{IEEEeqnarray*}
An immediate consequence of this definition is that the \mbox{$\leq$-relation} has the product as g.l.b., the addition as l.u.b., $1$ as top element and $0$ as bottom element.
Term~$1$ represents a value which holds by default, without an explicit cause, and will be assigned to the empty body.
Term~$0$ represents the absence of cause or the empty set of causes, and will be assigned to falsity.
Furthermore, applying distributivity (and absorption) of products and applications over addition, every term can be represented in a \emph{(minimal) disjunctive normal form} in which addition is not in the scope of any other operation and every pair of addends are pairwise $\leq$-incomparable.
This normal form emphasizes the intuition that addition~`$+$' separates alternative causes.
Consider, for instance, a program~\newprogram\label{prg:disjuntive.normal.form} containing the rule
\vspace{-0.25cm}
\begin{IEEEeqnarray}{l C ? r C l}
r_4 &:& \noheart &\lparrow& \dead
	\label{eq:r1.no-breathing}
\end{IEEEeqnarray}
plus program \programref{prg:billy}=$\{\eqref{eq:r1}, \eqref{eq:r2}, \eqref{eq:suzy}, \eqref{eq:r4a}, \eqref{eq:r4}\}$.
Program~\programref{prg:disjuntive.normal.form} has a least model that coincides with $I_{\ref{prg:billy}}$ of~\programref{prg:billy} in all atoms excepting for the new atom $\noheart$, whose justification becomes:
\begin{IEEEeqnarray}{l ? C ? c C c C l}
I_{\ref{prg:disjuntive.normal.form}}(\noheart)
	&=& \IEEEeqnarraymulticol{5}{c}{ I_{\ref{prg:disjuntive.normal.form}}(\dead) \cdot r_4 }
	\\
	&=& \big( \ (\loaded * \harvey \cdotl r_2) \cdotl r_1 &+& \billystab \cdotl r_3 \ \big)
		\cdot r_4
	\\
	&=&  (\loaded * \harvey \cdotl r_2) \cdotl r_1 \cdot r_4 &+& \billystab \cdotl r_3 \cdot r_4
	\label{eq:disjuntive.normal.form}
\end{IEEEeqnarray}
Expression~\eqref{eq:disjuntive.normal.form} is in minimal disjunctive normal and emphasizes the existence of two alternative causes.
In this sense, any term without addition $G \in \causes$ represents an individual cause, while terms with addition $t \in \values\backslash \causes$ will be used to represent sets of non-redundant causes.
\begin{figure}[htbp]\centering\footnotesize
$
\xymatrix @-4mm {
\\
  & {\loaded} \ar@/_13pt/[ddr]   &             & {\harvey} \ar[d]
  & \hspace{1.5cm} & {\billystab} \ar[dd]
  & \hspace{1.5cm} & {\billystab} \ar[dd] \ar@{.>}@(ul,ur)  \ar@/_20pt/@{.>}[ddd]\\
  &    &            & {r_2} \ar@/^/[dl]\\
  &                   & {r_1}\ar[d]  &
  & & {r_3} \ar[d]
  & & {r_3} \ar[d] \ar@{.>}@(ur,dr)\\
  &                   & {r_4}   &  
  & & {r_4}
  & & {r_4} \ar@{.>}@(dr,dl)\\
  &                   & {G_1}
  & & & {G_2}
  & & {G_2^*}
}
$
\caption{\small Graphs $G_1$ and $G_2$ representing the two alternative causes of $\noheart$ in program~\programref{prg:disjuntive.normal.form}.
Graph~$G_2^*$ is the reflexive and transitive closure of $G_2$ and, thus, the c-graph associated to~$G_2$.}
\label{fig:disjuntive.normal.form}
\end{figure}
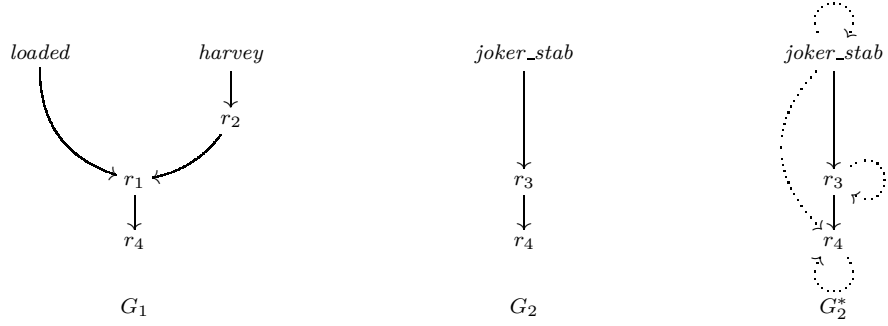

Furthermore, each individual cause $G \in \causes$ can be depicted as a graph whose vertices are labels.
For instance, graphs $G_1$ and $G_2$ in Figure~\ref{fig:disjuntive.normal.form} correspond to the two individual causes of the atom $\noheart$ with respect to the least model of~~\programref{prg:disjuntive.normal.form}.
This correspondence was formalised in~\cite{fandinno2015thesis} in the following way.

\begin{definition}[Causal graph]\label{def:causal.graph}
Given a set of labels $\lb$, a \emph{causal graph (or c-graph)}
$G \subseteq \lb \times \lb$ is a set of edges transitively and reflexively closed.
By $\graphs$ we denote the set of all c-graphs that can be formed with labels from $\lb$.
Furthermore, given two causal graphs $\set{G,G'} \subseteq \graphs$, by $G \cdot G'$ and $G * G'$, we denote the transitive closure of $G \cup G' \cup \setm{ (v,v') }{ (v,v) \in G \text{ and } (v',v') \in G' }$
and  $G \cup G'$, respectively.\qed
\end{definition}

\begin{theorem}[From~\protect\citeNP{fandinno2015thesis}]\label{thm:algebra.values}
The function \ \ $term: \graphs \longrightarrow \causes$ \ \ given by
\begin{gather*}
term(G) \ \ \mapsto \ \ \prod_{(v_1,v_2) \in G}  v_1 \cdotl v_2 
  \label{eq:algebra.values.isompphism}
\end{gather*}
is an isomorphism between algebras $\tuple{\graphs,*,\cdot,\emptyset,\subseteq}$ and $\tuple{\causes,*,\cdot,1,\leq}$.\QED
\end{theorem}

Defining causal graphs as transitively and reflexively closed is convenient for a simpler definition of product and application over graphs.
However, for the sake of readability, we will just depict a causal graph $G$ as one of its transitive and reflexive reductions\footnote{Recall that the transitive and reflexive reduction of a graph $G$ is a graph~$G'$ whose transitive and reflexive closure is~$G$.
A causal graph, in which every cycle is a reflexive edge, has a unique transitive and reflexive reduction.}.
For instance, Figure~\ref{fig:disjuntive.normal.form} shows a graph $G_2$ that will actually stand for its closure, the causal graph $G_2^*$.
Furthermore, we will also allow writing 
$I_{\ref{prg:disjuntive.normal.form}}(\noheart) = G_1 + G_2$ instead of~\eqref{eq:disjuntive.normal.form}.
It is also worth to mention that each path in a causal graph may be understood as causal chain in the sense of~\citeN{lewis1973causation}: a finite sequence of actual events $e_1,\dots,e_n$ where each $e_i$ causally depends on $e_{i-1}$ (for $0<i\leq n$).

\section{Causal stable models for disjunctive logic programs}
\label{sec:csm}

An \emph{interpretation} is a mapping \mbox{$\cI:At\longrightarrow\values$} assigning a value to each atom.
An interpretation~$I$ is said to be \emph{standard} or \emph{two-valued} when it maps each atom into the set~$\set{0,1}$.
The value assigned to a negative literal $\Not \rA$ by an interpretation $I$, denoted as
$I(\Not \rA)$, is  defined as: $I(\Not \rA) \,\eqdef\, 1$ if
\mbox{$I(\rA)=0$}; and $I(\Not \rA)\,\eqdef\, 0$ otherwise.
For any pair of interpretations $\cI$ and $\cJ$,
we write $\cI\leq \cJ$ to represent the straightforward causal ordering, that is, \mbox{$\cI(\rH) \leq \cJ(\rH)$} for every atom $\rA \in At$.


\REVIEW{R3.1}
Since body disjunction can be captured by addition~`$+$'~\cite{fandinno2015aspocp}, an immediate  attempt for interpreting disjunctive programs would consist in just defining
$I(\rA \vee \rA') = I(\rA) + I(\rA')$ and selecting $\leq$-minimal models.
However, this naive approach does not capture the intending meaning.
Consider for instance the following program~\newprogram\label{prg:naive}
\begin{gather*}
r_1 : \ \ head \vee tails \lparrow toss
\hspace{2cm}
a : \ \ toss
\hspace{2cm}
b : \ \ toss
\end{gather*}
where two persons $a$ and $b$ both give the command to $toss$ a coin.
One may expect that program~\programref{prg:naive}
has two causal stable models, one $I_{\ref{prg:naive}}$ in which 
$I_{\ref{prg:naive}}(head) = a \cdotl r_1^{head} + b \cdotl r_1^{head}$ and
$I_{\ref{prg:naive}}(tails) = 0$, and another $I_{\ref{prg:naive}}'$ in which
$I_{\ref{prg:naive}}'(head) = 0$ and
$I_{\ref{prg:naive}}'(tails) = a \cdotl r_1^{tails} + b \cdotl r_1^{tails}$.
However, 																	
under this naive definition,
there are other two additional causal stable stable models that unintendedly ``distribute'' the causes of $toss$ among $head$ and $tails$. We get, for instance, $I_{\ref{prg:naive}}''$ where 
$I_{\ref{prg:naive}}''(head) = a \cdotl r_1^{head}$ and
$I_{\ref{prg:naive}}''(tails) = b \cdotl r_1^{tails}$, and a dual model $I_{\ref{prg:naive}}'''$ that switches the roles of $a$ and $b$.

This example shows that, rather than taking the sum of all disjuncts in the head, the evaluation of rules should consider each head disjunct independently, as defined below.
\ENDREV

\begin{definition}[Model]\label{def:model}
An \emph{interpretation} $I$ satisfies rule $\R$ of the form of~\eqref{eq:rule} iff
\begin{eqnarray}
\big( \, I(\rB_1) * \dotsc * I(\rB_\npbody) 
	* I(\Not \rB_{\npbody+1})* \dotsc * I(\Not \rB_\nnbody)
		\, \big) \cdot r_i \cdot \rH_j \ \ \leq \ \ I(\rH_j)
\end{eqnarray}
for some atom $\rH_j \in head(\R)$.
We say that an interpretation~$I$ is a \emph{model} of a program~$\cP$, in symbols $I \models P$, iff $I$ satisfies all rules in~$\cP$.
Moreover, if $I$ is standard (two-valued) we further write $I \stmodels P$ and call it \emph{standard model}.\qed
\end{definition}

\begin{observation}{\label{prop:}}
If $r$ is a fact $A$, that is, it has the form $(A : A \lparrow)$ then $I \models r$ iff $I(A) \geq A \cdot A = A$ (by idempotence of application on labels).\qed
\end{observation}

\REVIEW{R3.8}
It is easy to see that, under Definition~\ref{def:model},
interpretations
$I_{\ref{prg:naive}}''$
and
$I_{\ref{prg:naive}}'''$
are no longer models of~\programref{prg:naive}.
Still, a second issue comes from selecting $\leq$-minimal models.
Consider, for instance, the following program
\newprogram\label{prg:naive2}
\begin{gather*}
r_1 : \ \ head \vee tails
\hspace{2cm} head
\end{gather*}
which has two $\leq$-minimal models, one in which $I_{\ref{prg:naive2}}(head) = head + r_1^{head}$ and $I_{\ref{prg:naive2}}(tails) = 0$, plus another in which $I_{\ref{prg:naive2}}'(head) = head$ and
$I_{\ref{prg:naive2}}'(tails) = r_1^{tails}$.
However, only the former corresponds to the unique GL-stable model of this program, which leaves $tails$ false under its truth-minimality criterion. To overcome this issue, we will define an extra partial order~$\wleq$.
First, for any interpretation~$I$ we define its corresponding standard interpretation, $I^{st}$, as follows:
\begin{gather*}
I^{st}(\rH) \ \ \eqdef \ \ \begin{cases}
        1 &\text{iff } I(\rH) > 0\\
        0 &\text{iff } I(\rH) = 0
\end{cases}
\end{gather*}
We define the set of true atoms in an interpretation $I$ as $\Atoms(I)\eqdef \setm{A \in At}{I(A) \neq 0}$.%
\ENDREV%
\begin{Proposition}{\label{prop:positive.standard.models}}
For any standard interpretation $I$ and positive program $P$: $I \stmodels P$ iff $\Atoms(I)$ is closed under $P$.\qed
\end{Proposition}
Then, for any pair of interpretations $\cI$ and $\cJ$,
we write $\cI \wleq \cJ$ when either $\cI \leq \cJ$ or $\Atoms(I) \subset \Atoms(J)$.
That is, $\cI \wleq \cJ$ is a weaker relation, since apart from the cases in which $\cI \leq \cJ$ it also holds when true atoms in $I$ are a strict subset of true atoms in~$J$.
As usual, we write $I<J$ (resp. $I \wless J$) iff $I\leq J$ (resp $I\wleq J$) and $I\neq J$.
Notice that $Atoms(I) \subset Atoms(J)$ implies $I\neq J$ and so $I \wless J$.
We say that an interpretation $I$ is $\leq$-minimal (resp. $\wleq$-minimal) satisfying some property when there is no $J < I$ (resp. $J \wless I)$ satisfying that property.
It is worth to notice that there is a \mbox{$\leq$-bottom} and \mbox{$\wleq$-bottom} interpretation~\botI\ (resp. a $\leq$-top and $\wleq$-top interpretation~\topI) that stands for the interpretation mapping every atom~$\rA$ to $0$ (resp. $1$).


\begin{definition}[Causal/standard stable model of a positive program]
Let $P$ be a positive program.
A model~$I$ of~$P$ is a \emph{causal stable} model iff it is \mbox{$\wleq$-minimal} among models of~$P$.
A standard model~$I$ of~$P$ is a \emph{standard stable} model iff $I$ is \mbox{$\wleq$-minimal} among standard models of $P$.\QED
\end{definition}

Standard stable models have a straightforward relation to GL-stable models, as stated below:

\begin{Theorem}{\label{thm:positive.standard<->standard.semantics}}
Let $P$ be a positive program.
An interpretation $I$ is a standard stable model of $P$ iff $\Atoms(I)$ is a GL-stable model of~$P$.\qed
\end{Theorem}
From now on we use both concepts indistinctly.
Note that not any standard stable model $I$ of~$P$ needs to be causal stable, since there could exist another non-standard model $J \wless I$.
Still, there exists a strong connection between standard and causal stable models.


\begin{Theorem}{\label{thm:positive.standard<->causal.smodel}}
Let $P$ be a positive program.
Then, a standard interpretation $J$ is a standard stable model of $P$ iff there is some causal stable model $I$ of $P$ such that $I^{st} = J$.\qed
\end{Theorem}

In other words, there is a one-to-one correspondence between each standard stable model and a class of causal stable models that agree in the truth assignment of atoms (but may vary in their explanations).
For instance,~\programref{prg:coin}=$\{\eqref{eq:r1},\eqref{eq:suzy},\eqref{eq:r2b},\eqref{eq:r3}\}$ has two GL-stable models,
$S_{\ref{prg:coin}} = \set{\harvey, \heads}$ and $S_{\ref{prg:coin}}' = \set{\harvey, \tails, \shoot, \dead}$.
Notice that any model~$I_{\ref{prg:coin}}$ of program~\programref{prg:coin} must satisfy
$I_{\ref{prg:coin}}(\harvey) \geq \harvey$ and, thus, any $\wleq$-minimal model must satisfy that
$I_{\ref{prg:coin}}(\harvey) = \harvey$.
Similarly, any model $I_{\ref{prg:coin}}$ must satisfy either
$I_{\ref{prg:coin}}(\heads) \geq \harvey \cdotl r_1 \cdotl \heads$
or
$I_{\ref{prg:coin}}(\tails) \geq \harvey \cdotl r_1 \cdotl \tails$
and, thus,
there are two \mbox{$\wleq$-minimal} models, $I_{\ref{prg:coin}}$ and $I'_{\ref{prg:coin}}$, of~\programref{prg:coin} satisfying:
\begin{gather*}
\begin{IEEEeqnarraybox}[][t]{l ; C ; l}
I_{\ref{prg:coin}}(\harvey) &=& \harvey
\\
I_{\ref{prg:coin}}(\heads) &=& \harvey \cdotl r_1 \cdotl \heads
\\
I_{\ref{prg:coin}}(\tails) &=& 0
\\
I_{\ref{prg:coin}}(\shoot) &=& 0
\\
I_{\ref{prg:coin}}(dead) &=& 0
\end{IEEEeqnarraybox}
\hspace{1.5cm}
\begin{IEEEeqnarraybox}[][t]{l ; C ; l}
I_{\ref{prg:coin}}'(\harvey) &=& \harvey
\\
I_{\ref{prg:coin}}'(\heads) &=& 0
\\
I_{\ref{prg:coin}}'(\tails) &=& \harvey \cdotl r_1 \cdotl \tails
\\
I_{\ref{prg:coin}}(\shoot) &=& \harvey \cdotl r_1 \cdotl \tails \cdotl r_2 \cdotl \shoot
\\
I_{\ref{prg:coin}}(\dead) &=& \harvey \cdotl r_1 \cdotl \tails \cdotl r_2 \cdotl \shoot \cdotl r_3 \cdotl  \dead
\end{IEEEeqnarraybox}
\end{gather*}
For the sake of clarity, we will write $r_i^\rA$ instead of $r_i \cdotl \rA$ when \mbox{$r_i \in \lbr$} and $\rA \in \at$.
Intuitively, symbol $r_i^\rA$ means that among the possible consequences of rule $r_i$, it has caused~$\rA$.
For normal rules like $r_1$ and $r_2$ stating the selected atom is redundant and, thus, we will usually omit the head atom label.
Using these conventions, we may rewrite the $\wleq$-minimal models of~\programref{prg:coin} as
\begin{gather*}
\begin{IEEEeqnarraybox}[][t]{l ; C ; l}
I_{\ref{prg:coin}}(\harvey) &=& \harvey
\\
I_{\ref{prg:coin}}(\heads) &=& \harvey \cdotl r_1^{\heads}
\\
I_{\ref{prg:coin}}(\tails) &=& 0
\\
I_{\ref{prg:coin}}(\shoot) &=& 0
\\
I_{\ref{prg:coin}}(\dead) &=& 0
\end{IEEEeqnarraybox}
\hspace{2.25cm}
\begin{IEEEeqnarraybox}[][t]{l ; C ; l}
I_{\ref{prg:coin}}'(\harvey) &=& \harvey
\\
I_{\ref{prg:coin}}'(\heads) &=& 0
\\
I_{\ref{prg:coin}}'(\tails) &=& \harvey \cdotl r_1^{\tails}
\\
I_{\ref{prg:coin}}'(\shoot) &=& \harvey \cdotl r_1^{\tails} \cdotl r_2 \\
I_{\ref{prg:coin}}'(\dead) &=& \harvey \cdotl r_1^{\tails} \cdotl r_2 \cdotl r_3
\end{IEEEeqnarraybox}
\hspace{.5cm}
\end{gather*}
Note that 
$I'_{\ref{prg:coin}}(\dead)$ is just the justification of $\dead$ mentioned in the introduction~\eqref{eq:introduction.term}.
Another interesting observation is that the sets of true atoms in each model, $\Atoms(I_{\ref{prg:coin}})=\set{\harvey, \heads}$ and $\Atoms(I_{\ref{prg:coin}}')=\set{\harvey, \tails, \shoot, \dead }$, respectively corresponding to the two GL-stable models of~\programref{prg:coin}.
This correspondence, however, is not always one-to-one, as stated by Theorem~\ref{thm:positive.standard<->causal.smodel}.
To see why, consider a program~\newprogram\label{prg:non.one-to-one} consisting of the following rules
\begin{gather*}
r_1 : \ \ a \veeh b \lparrow
\hspace{2cm}
r_2 : \ \ a  \lparrow b
\hspace{2cm}
r_3 : \ \ b \lparrow a
\end{gather*}
Program~\programref{prg:non.one-to-one} has a unique GL-stable model~$\set{a, b}$ but two causal stable models $I_{\ref{prg:non.one-to-one}}$ and $I'_{\ref{prg:non.one-to-one}}$ such that:
\vspace{-0.25cm}
\begin{gather*}
\begin{IEEEeqnarraybox}[][t]{l ; C ; l}
I_{\ref{prg:non.one-to-one}}(a) &=& r_1^a
\\
I_{\ref{prg:non.one-to-one}}(b) &=& r_1^a \cdotl r_3
\end{IEEEeqnarraybox}
\hspace{3cm}
\begin{IEEEeqnarraybox}[][t]{l ; C ; l}
I'_{\ref{prg:non.one-to-one}}(a) &=& r_1^b \cdotl r_2
\\
I'_{\ref{prg:non.one-to-one}}(b) &=& r_1^b
\end{IEEEeqnarraybox}
\end{gather*}
As we can see, the true atoms in both models $Atoms(I)=Atoms(I')=\set{a,b}$ coincide with the unique GL-stable model, but their \emph{explanations differ}.
In $I_{\ref{prg:non.one-to-one}}$, atom $a$ is a (non-deterministic) effect of the disjunction $r_1$, while $b$ is derived from $a$ through $r_3$.
Analogously, $I'_{\ref{prg:non.one-to-one}}$ makes $b$ true because of $r_1$ and then obtains $a$ from $b$ through $r_2$.

\paragraph{Programs with negation.}
To introduce default negation, let us consider a variation of our running example in which shooting the victim may fail in several ways: the victim may be wearing a $\bulletproof$ vest, the gunpowder may be $\wet$, etc.
A possible encoding of this scenario is program~\newprogram\label{prg:shoot.fail}
obtained from program~\programref{prg:coin}=$\{\eqref{eq:r1},\eqref{eq:suzy},\eqref{eq:r2b},\eqref{eq:r3}\}$ after replacing rule~\eqref{eq:r1} by
\vspace{-0.25cm}
\begin{IEEEeqnarray}{l C ? r C l}
r_1 &:& \dead &\lparrow& \shoot, \Not \ab
	\label{eq:r1.not}
\end{IEEEeqnarray}
and rules for every possible exception in the following way
\begin{IEEEeqnarray}{l C ? r C l}
r_5 &:& \ab &\lparrow& \wet
	\label{eq:ab.bulletproof}
\\
r_6 &:& \ab &\lparrow& \bulletproof
	\label{eq:ab.wet}
\end{IEEEeqnarray}
From a causal perspective, saying that the lack of an exception is part of a cause (e.g., for $\dead$) may seem rather counterintuitive.
It is not the case that we are $dead$ because the gunpowder was not $\wet$,
or because we are not wearing a $\bulletproof$ vest,
or whatever other possible exception that might be added in the future\footnote{A case of the well-known \emph{qualification problem}~\cite{mccarthy1987epistemological}, i.e., the impossibility of listing all the possible conditions that prevent an action to cause a given effect.}\!\!.\,
Instead, as nothing violated default~\eqref{eq:r1.not},
the justifications for $\dead$ should be the same as in  program~\programref{prg:coin}.
In this sense, falsity in understood as the \emph{default situation} that is broken when a cause is found\footnote{The paper~\cite{hitchcock2009cause} contains an extended discussion with several examples showing how people ordinarily understand causes as deviations from a norm.}\!\!.\, This interpretation carries over to negative literals, so that the presence of $\Not \rA$ in a rule body does not propagate causal information, but instead, it simply checks the absence of an exception. To capture this behaviour, we proceed to extend the traditional program reduct~\cite{GelfondL88} to causal logic programs.

\begin{definition}[Program reduct]\label{def:reduct}
The \emph{reduct} of a program $P$ with respect to an interpretation $I$, in symbols~$P^I$, is the result of
\begin{enumerate}
\item removing all rules such that $I(\rB) \neq 0$ for some $\rB \in \body^-(\R)$,
\item removing all rules such that $I(\rB) = 0$ for some $\rB \in \head^-(\R)$,
\item removing all the negative literals for the remaining rules.\qed
\end{enumerate}
\end{definition}

\begin{definition}[Causal stable model]\label{def:causal.smodel}
We say that an interpretation $I$ is a \emph{causal stable model} of some program~$P$ iff $I$~is a causal stable model of the positive program~$P^I$.
\end{definition}

\begin{observation}\label{obs:reduct}
Note that $P^I$ coincides with $P^{I^{st}}$, i.e., the reduct does not vary if we just use the standard interpretation $I^{st}$. Moreover, it also coincides with the classical GL-reduct $P^{\Atoms(I)}$.\qed
\end{observation}

\begin{Theorem}{\label{thm:standard<->causal.smodel}}
Let $P$ be a labelled program.
Then, $J$ is a standard stable model (i.e. $\Atoms(J)$ is a GL-stable model) of $P$ iff there is some causal stable model $I$ of $P$ such that $I^{st} = J$ (i.e. $Atoms(I)=Atoms(J)$).\qed
\end{Theorem}

Theorem~\ref{thm:standard<->causal.smodel} is an immediate consequence of
Observation~\ref{obs:reduct} plus Theorem~\ref{thm:positive.standard<->causal.smodel}.
Furthermore, it is easy to see that program~\programref{prg:shoot.fail} has the same two standard stable models of~\programref{prg:coin}.
The reduct of~\programref{prg:shoot.fail} with respect to interpretations~$I_{\ref{prg:coin}}$ and~$I_{\ref{prg:coin}}'$
is just the result of removing $\Not \ab$ from the body of~\eqref{eq:r1.not}.
Thus, these are also the two causal stable models of~\programref{prg:shoot.fail}.
On the other hand, a program~\newprogram\label{prg:shoot.fail.wet}, obtained by adding the fact~$\wet$ to program~\programref{prg:shoot.fail},
has two standard stable models
$\set{\harvey, \heads, \wet}$ and
$\set{\harvey,  \tails, \wet}$ and two corresponding causal stable models.
Note that the reduct of program~\programref{prg:shoot.fail} w.r.t.
$I_{\ref{prg:shoot.fail}}$
and
$I_{\ref{prg:shoot.fail}}'$
is just the result of removing rule~\eqref{eq:r1.not} and, thus, there is no justification for $\dead$ in any of these two causal stable models.


\section{Causal-choice rules}
\label{sec:causal-choice.rules}

Disjunctive rules are useful for representing non-deterministic causal laws that represent the possible outcomes of an experiment like throwing a coin.
However, disjunction alone is not enough to capture the causal nature of some non-deterministic statements of the form ``$A$ may cause $B$.''
In order to illustrate this limitation, consider the statement 
``An $\infection$ may cause the patient to have $\fever$.''
A frequent way to represent this kind of statements in ASP is using a choice rule as follows
\begin{IEEEeqnarray}{l C ? r C l C l}
r_1 &:& \fever &\veeh& \Not \ \fever &\lparrow& \infection
	\label{eq:fever.infection}
\\
	&& && \infection
	\label{eq:infection}
\end{IEEEeqnarray}
Program~\newprogram\label{prg:infection} consisting of rules~(\ref{eq:fever.infection}-\ref{eq:infection}) has two standard stable models
$S_{\ref{prg:infection}} =\set{\fever}$ and
$S_{\ref{prg:infection}}' =\set{\infection, \fever}$ and, in fact, two respective causal stable models $I_{\ref{prg:infection}}$ and $I_{\ref{prg:infection}}'$:
\begin{gather*}
\begin{IEEEeqnarraybox}[][t]{l ; C ; l}
I_{\ref{prg:infection}}(\infection)   &=&  \infection\\
I_{\ref{prg:infection}}(\fever) &=& 0
\end{IEEEeqnarraybox}
\hspace{2cm}
\begin{IEEEeqnarraybox}[][t]{l ; C ; l}
I_{\ref{prg:infection}}'(\infection)   &=&  \infection\\
I_{\ref{prg:infection}}'(\fever) &=& \infection \cdotl r_1^{\fever}
\end{IEEEeqnarraybox}
\end{gather*}
which obtain the expected result.
However, the causal behaviour of~\eqref{eq:fever.infection} is not exactly what we look for but, instead, would correspond to the assertion ``$\infection$ may cause $\fever$ or $\fever$ is false.''
This becomes evident if we add a second external cause for $\fever$:
\begin{IEEEeqnarray}{l C ? r C l}
r_2 &:& \fever &\lparrow&
	\label{eq:fever}
\end{IEEEeqnarray}
The unique standard stable model of the new program~\newprogram\label{prg:infection.two} formed by~\programref{prg:infection} plus~\eqref{eq:fever} is now~$S_{\ref{prg:infection}}'$, since once $\fever$ is fixed by $r_2$, the choice $r_1$ \emph{is forced to cause} $\fever$. 
This ``forced causation'' is irrelevant when we do not consider causal explanations: in fact, under \mbox{GL-stable} models semantics, once we add a fact for $\fever$, rule~$r_1$ becomes a tautology and can always be safely removed from the program.
However, $r_1$ cannot be removed under the causal semantics (since it expresses some relation between $\infection$ and $\fever$) and, as \eqref{eq:fever} fixes $\fever$ true for any model $I$, the corresponding reduct of $r_1$ will mandatorily be:
\begin{IEEEeqnarray}{l C ? r C l}
r_1 &:& \fever  &\lparrow& \infection
	\label{eq:fever.infection.reduced}
\end{IEEEeqnarray}
Under these circumstances, it is easy to see that we get a \emph{unique} causal stable model~$I_{\ref{prg:infection.two}}$ of program~\programref{prg:infection.two} that satisfies $I_{\ref{prg:infection.two}}(\fever) = \infection \cdotl r_1^{\fever} + r_2$.
In other words, we get now that $\infection$ is always forced to be one of the causes of $\fever$, although the only addition we made was providing an (independent) alternative cause $r_2$.

For capturing the correct meaning of ``may cause'' in this example, we should have obtained a second causal stable model $I'_{\ref{prg:infection.two}}$ where $I'_{\ref{prg:infection.two}}(\fever) = r_2$, that is, $\fever$ is still true due to $r_2$ but the $\infection$ does not result to be an alternative cause of $\fever$ this time.
The problem with this second model is that while $\Atoms(I'_{\ref{prg:infection.two}})=\Atoms(I_{\ref{prg:infection.two}})$ means that the reduct for both models will always coincide, and we also have\footnote{Indeed, $I'_{\ref{prg:infection.two}}(\fever) = r_2 < \infection \cdotl r_1^{\fever} + r_2 = I_{\ref{prg:infection.two}}(\fever)$ while $I_{\ref{prg:infection.two}}(\infection)=I'_{\ref{prg:infection.two}}(\infection)=\infection$.} 
$I'_{\ref{prg:infection.two}} \wless I_{\ref{prg:infection.two}}$, meaning that no program may have \emph{both} interpretations as $\wleq$-minimal models of such a reduct. Hence, it is clear that no program using the previous language can capture the intended models of this example.
We introduce next the notion of \emph{causal-choice rule} (and its associated program reduct) that intuitively captures the idea of ``$A$ may cause $B$'' and that, when causal information is disregarded, amounts to a usual choice rule under the standard stable model semantics.

\begin{definition}\label{def:causal.choice}
A \emph{causal-choice rule} is an expression of the form of
\begin{eqnarray}
r_i: \ \ \rH_1 \veeh \dotsc \veeh \rH_\nphead
 \veeh \Not \rH_{\nphead + 1} \veeh \dotsc \veeh \Not \rH_\nnhead
		\ \warrow \ \rB_1, \dotsc, \rB_\npbody,
		\Not \rB_{\npbody+1}, \dotsc, \Not \rB_\nnbody
    \label{eq:choice.rule}
\end{eqnarray}
where \mbox{$r_i\in \lb$} is a label and all $\rA_j \in \at$ and $\rB_j$ are atoms.
For every causal-choice rule~$\R$ of the form of~\eqref{eq:choice.rule}, by $rule(\R)$ we denote its corresponding rule of the form of~\eqref{eq:rule}, i.e., $rule(\R)$ corresponds to replacing `$\warrow\!$' by `$\lparrow$.'\qed
\end{definition}

\begin{definition}\label{def:reduct.ext}
The reduct of a program with causal-choice rules~$P$, with respect to an interpretation $I$, in symbols also $P^I$, is obtained by
\begin{enumerate}
\item removing every causal-choice rule $\R$ such that $I$ does not satisfy~$rule(\R)$,
\item replacing the remaining causal-choice rules $\R$ by~$rule(\R)$
\item apply the reduct of Definition~\ref{def:reduct} to the result of the previous two steps.
\end{enumerate}
An interpretation $I$ is a causal stable model of a program with causal-choice rules~$P$ iff $I$ is a causal stable model of $P^I$.\qed
\end{definition}

We can now represent the variation of our running example with program~\newprogram\label{prg:infection.choice}:
\begin{IEEEeqnarray}{l C ? c C l}
r_1 &:& \fever &\warrow& \infection
	\label{eq:fever.infection.choice}
\\
r_2 &:& \fever &\lparrow&
	\label{eq:fever.2}
\\
	&& \infection
	\label{eq:infection.2}
\end{IEEEeqnarray}
On the one hand, the reduct of program~\programref{prg:infection.choice} w.r.t.
$I_{\ref{prg:infection.two}}$
is the program obtained by replacing the choice rule~\eqref{eq:fever.infection.choice} by its corresponding rule~\eqref{eq:fever.infection.reduced}.
It is easy to check that
$I_{\ref{prg:infection.two}}$ is a $\wleq$-minimal model of~$\programref{prg:infection.choice}^{I_{\ref{prg:infection.two}}}$ and, thus, a causal stable model of $\programref{prg:infection.choice}$.
On the other hand,
the reduct of  program~\programref{prg:infection.choice} w.r.t.
$I_{\ref{prg:infection.two}}'$
is the result of removing rule~\eqref{eq:fever.infection.choice}
and, thus,
$I_{\ref{prg:infection.two}}'$ is also a $\wleq$-minimal model of~$\programref{prg:infection.choice}^{I_{\ref{prg:infection.two}}'}$ and a causal stable model of program~$\programref{prg:infection.choice}$.

As explained before, this example illustrates that there are programs with causal-choice rules whose causal stable models cannot be captured by programs without them.
Despite that, it can be shown that, when causal information is disregarded, causal-choice rules behave like standard choice rules.
The following definition and Theorem~\ref{thm:reduct.with.choices.correspondence} below formalise this intuitive idea.

\begin{definition}
For any program $P$ and causal-choice rule $\R \in P$ of the form of~\eqref{eq:choice.rule},
by $\rchoice{\R}{P}$, we denote the program obtained after replacing $\R$ by a set of rules of the form~of
\begin{eqnarray}
r_i: \ \rH_1 \!\veeh\! \dotsc \!\veeh\! \rH_j \!\veeh\! \Not \rH_j \!\veeh\! \dotsc \!\veeh\! \rH_\nphead
 \!\veeh\! \Not \rH_{\nphead + 1} \!\veeh\! \dotsc \!\veeh\! \Not \rH_\nnhead
		\lparrow body(\R)
    \label{eq:choice.rule.remove}
\end{eqnarray}
per each atom~$\rA_j \in \head^+(\R)$.\qed
\end{definition}

As an example, the causal-choice~\eqref{eq:fever.infection.choice} in a program $P$ would be replaced by the standard choice~\eqref{eq:fever.infection} in $\rchoice{\eqref{eq:fever.infection.choice}}{P}$.

\begin{Theorem}{\label{thm:reduct.with.choices.correspondence}}
Let $P$ be a program with causal-choice rules,
$\R \in P$ be a causal-choice rule 
and let~$I$ be a standard interpretation.
Then, $I$ is a standard stable model of $P$ iff $I$ is a standard stable model of $\rchoice{\R}{P}$.\qed
\end{Theorem}

\section{Related Work} \label{sec:related}

The most obvious related work is our previous approach for non-disjunctive programs~\cite{CabalarFF14}.
The following result established a one-to-one correspondence between our causal stable models and the \emph{CJ-stable models} from~\cite{CabalarFF14}.

\begin{Theorem}{\label{thm:old.correspondence}}
Let $P$ be a normal program, 
for every causal stable model $I$,
there is a unique CJ-causal stable model $J$ such that
$I^{st} = J^{st}$, and vice-versa.
Furthermore, if $I$ and $J$ are respectively a causal stable model and CJ-stable model of~$P$ such that $I^{st} = J^{st}$,
then
there is a justification $G' = \noatomlabels{G}$ such that $G' \leq J(\rA)$
for every atom~$\rA$ and justification $G \leq I(\rA)$,
where $\noatomlabels{G}$ denotes the result of removing every edge containing an atom label $\rA \in \at$ in the causal graph~$G$.
\qed
\end{Theorem}

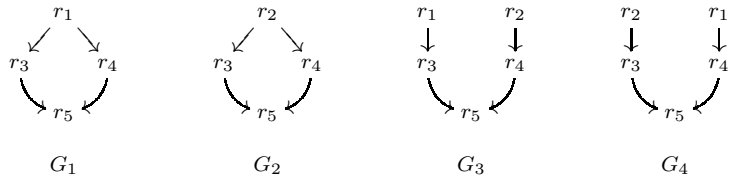
\begin{figure}[hbtp]\centering\footnotesize
$
\xymatrix @R3mm @C-8mm{
					& {r_1} \ar[dl] \ar[dr]		&							&  \hspace{0.75cm} &
					& {r_2} \ar[dl] \ar[dr]		&							&  \hspace{0.75cm} &
	{r_1} \ar[d]		&   	    				& {r_2} \ar[d]				&  \hspace{0.75cm} &
	{r_2} \ar[d]		&	       					& {r_1} \ar[d]				\\
	{r_3} \ar@/_/[dr] 	&             				& {r_4} \ar@/^/[dl]			&			&
	{r_3} \ar@/_/[dr] 	&             				& {r_4} \ar@/^/[dl]			&			&
	{r_3} \ar@/_/[dr] 	&             				& {r_4} \ar@/^/[dl]			&			&
	{r_3} \ar@/_/[dr] 	&             				& {r_4} \ar@/^/[dl]			\\
	                	& {r_5}   				&                 			&			&
	                	& {r_5}   				&                 			&			& 
	                	& {r_5}   				&                 			&			& 
	                	& {r_5}   				&                 			\\
%
%
%
%
	                	& {G_1}       				&                 			&			&
	                	& {G_2}       				&                 			&			&
	                	& {G_3}       				&                 			&			&
	                	& {G_4}       				&                 			
}
$
\caption{Causal graphs associated with $d$ by the unique CJ-stable model of program~\programref{prg:old.difference}.}
\label{fig:graphs.no-atom-labelling}
\end{figure}

Note, that causal justifications with respect to any causal stable model are always a (possibly strict) subset of the causal justifications with the corresponding CJ-stable model.
To illustrate this difference, consider
the following normal program~\newprogram\label{prg:old.difference}
\begin{gather*}
\begin{IEEEeqnarraybox}[][t]{l C ? r C l C l}
r_1 &:& a &\lparrow& 
\end{IEEEeqnarraybox}
\hspace{1cm}
\begin{IEEEeqnarraybox}[][t]{l C ? r C l C l}
r_3 &:& b &\lparrow& a
\end{IEEEeqnarraybox}
\hspace{1cm}
\begin{IEEEeqnarraybox}[][t]{l C ? r C l C l}
r_4 &:& c &\lparrow& a
\end{IEEEeqnarraybox}
\hspace{1cm}
\begin{IEEEeqnarraybox}[][t]{l C ? r C l C l}
r_5 &:& d &\lparrow& b, c
\end{IEEEeqnarraybox}
\end{gather*}
whose unique CJ-stable model satisfies
$I_{\ref{prg:old.difference}}(d) \, = \, G_1$
with each $G_1$ being the causal graph~$G_1$ depicted in
Figure~\ref{fig:graphs.no-atom-labelling}.
Similarly, the unique causal stable model of program~\programref{prg:old.difference} satisfies
$I_{\ref{prg:old.difference}}(d) \ = \ G_1'$
where $G_1'$ is the result of removing every edge containing an atom label $\rA \in \at$ in the causal graph~$G_1$.
However,
if we consider a program~\newprogram\label{prg:old.difference.2} obtained by adding rule $(r_2 : a \lparrow )$ to program~\programref{prg:old.difference},
then
this program has a unique CJ-stable model that satisfies
$J_{\ref{prg:old.difference.2}}(d) \ = \ G_1 + G_2 + G_3 + G_4$ with each $G_j$ the corresponding causal graph depicted in Figure~\ref{fig:graphs.no-atom-labelling}.
On the other hand, under our current definition, 
program~\programref{prg:old.difference.2} has a unique causal stable model satisfying
$I_{\ref{prg:old.difference.2}}(d) \ = \ G_1' + G_2'$ with $G_1$ and $G_2$ being the result of removing every edge containing an atom label $\rA \in \at$ in the causal graphs~$G_1'$ and $G_2'$, respectively.
Note that causal graphs $G_3$ and $G_4$ can be considered somehow redundant with respect to $G_1$ and $G_2$ and, in this sense, our current definition, not only extends the definition of causal stable models given in~\cite{CabalarFF14} for disjunctive logic programs, but also better captures the notion of non-redundant justification for non-disjunctive programs (for more details we refer to~\citeNP{CabalarF16arXiv}).

Papers on reasoning about actions and change~\cite{Lin95,mccain1997causality,Thielscher97} have been traditionally focused on using causal inference to solve representational problems (mostly, the frame, ramification and qualification problems) without paying much attention to the derivation of cause-effect relations.
Perhaps the most established AI approach for causality is relying on \emph{causal networks}~\cite{Pearl00}.
In this approach, it is possible to conclude cause-effect relations like ``$A$ has caused $B$'' from the behaviour of structural equations by applying the counterfactual interpretation: ``had $A$ not happened, $B$ would not have happened.''
As discussed by~\citeN{hall2004}, this counterfactual-based definition corresponds to recognising some kind of \emph{dependence} relation in the behaviour of a non-causal system description. As opposed to this, Hall considers a different (and incompatible) definition where causes must be connected to their effects via \emph{sequences of causal intermediates}, something that is closer to our explanations in terms of causal graphs.
A similar approach has also been studied by~\citeN{Vennekens11} in CP-logic.


Focusing on LP, our work obviously relates to approaches on
justifications~\cite{specht1993generating,pemmasani2004online,pontelli2009justifications,damasio2013justifications,deneckerBS15,schulzT2016justifying}
which, as mentioned in the introduction, are limited to non-disjunctive programs.
Among these,
Why-not Provenance Justifications (WnP)~\cite{damasio2013justifications} \review{R3.10}{share with our approach} a semantic definition in terms of algebraic operations.
A formal comparison was done in~\cite{CabalarFF14inhibitors}.
With respect to~\cite{pontelli2009justifications}, a formal relation has not been established yet.
We conjecture that our causal explanations can be seen as non-redundant off-line justifications with respect to a kind of ``maximal set of assumptions.''
Still, this would apply to non-disjunctive programs only, since~\cite{pontelli2009justifications} do not cover disjunction.


A more far-fetched resemblance exists with respect to other meta-programming techniques for ASP debugging~\cite{GPST08,oetsch2010} which have a different goal, tackling the question of why some interpretations are not stable models of a (possibly disjunctive) program without negative literals in the head.
These approaches can be used to generate explanations for a set of literals $L$, with respect to an interpretation~$I$, by recursively asking why $I\backslash L$ is not an stable model of the program.
However, the given explanation cannot be used for our purpose of identifying causal relations, since it just provides a plain set of rules, without distinguishing which of them are a part of the same justification or which are their dependences.

\section{Conclusions and open issues} \label{sec:conc}

We have provided an extension of a logic programming semantics based on causal justifications to cope with disjunctive programs.
As in the previous, non-disjunctive approach, each true atom is assigned a set of non-redundant justifications, expressions built with rule labels and three algebraic operations: addition, product and application.
We have shown that, for each standard stable model of a program, there is a class of causal stable models that capture the different ways in which (the same) true atoms can be justified.
We have also shown that disjunctive rules are not enough to capture informal statements like ``$A$ may cause $B$'' and introduced causal-choice rules to represent them.

Regarding complexity, since the existence of causal stable models is completely determined by the existence of standard stable models, it intermediately follows that this problem is \mbox{$\SigmaP{2}$-complete} \cite{eiter1995computational}.
For query answering, it has been shown in~\cite{CabalarFF14Jelia}, that deciding whether a causal graph is a brave necessary cause is $\SigmaP{2}$-complete, even in the non-disjunctive case.
Whether disjunction makes the problem harder is still an open question.

Several other topics remain open for future study.
Interesting topics include a complexity assessment or studying an extension to arbitrary theories as with Equilibrium Logic \cite{Pea06} for the non-causal case. 
Further ongoing work is focused on implementation, the introduction of strong negation, a formal comparative with~\cite{pontelli2009justifications} and~\cite{schulzT2016justifying}.

\bibliographystyle{acmtrans}
\bibliography{refs}

\end{document}